\documentclass[twocolumn,english,aps,prl,floatfix,groupedaddress,showpacs,superscriptaddress]{revtex4}
\usepackage{amsmath}
\usepackage{amsbsy}
\usepackage{graphicx}
\usepackage{units}
\usepackage{xcolor} 
\usepackage{soul} 
\usepackage{amsmath}
\usepackage{amsfonts}
\usepackage{mathrsfs}
\usepackage{amsbsy}
\usepackage{bm}
\usepackage{babel}

\newcommand{\hh}{\ensuremath{H}}
\newcommand{\kk}{\ensuremath{K}}

\begin{document}

\title{Magnonlike Dispersion of Spin Resonance in Ni-doped BaFe$_{2}$As$_2$}

\author{M.~G.~Kim}\thanks{Current electronic address: mgkim@lbl.gov}
\affiliation{Ames Laboratory and Department of Physics and Astronomy, Iowa State University, Ames, IA, 50011, USA}
\author{G.~S.~Tucker}
\affiliation{Ames Laboratory and Department of Physics and Astronomy, Iowa State University, Ames, IA, 50011, USA}
\author{D.~K.~Pratt}
\affiliation{Ames Laboratory and Department of Physics and Astronomy, Iowa State University, Ames, IA, 50011, USA}
\author{S.~Ran}
\affiliation{Ames Laboratory and Department of Physics and Astronomy, Iowa State University, Ames, IA, 50011, USA}
\author{A.~Thaler}
\affiliation{Ames Laboratory and Department of Physics and Astronomy, Iowa State University, Ames, IA, 50011, USA}
\author{A.~D.~Christianson}
\affiliation{Oak Ridge National Laboratory, Oak Ridge, TN, 37831, USA}
\author{K.~Marty}
\affiliation{Oak Ridge National Laboratory, Oak Ridge, TN, 37831, USA}
\author{S.~Calder}
\affiliation{Oak Ridge National Laboratory, Oak Ridge, TN, 37831, USA}
\author{A.~Podlesnyak}
\affiliation{Oak Ridge National Laboratory, Oak Ridge, TN, 37831, USA}
\author{S.~L.~Bud'ko}
\affiliation{Ames Laboratory and Department of Physics and Astronomy, Iowa State University, Ames, IA, 50011, USA}
\author{P.~C.~Canfield}
\affiliation{Ames Laboratory and Department of Physics and Astronomy, Iowa State University, Ames, IA, 50011, USA}
\author{A.~Kreyssig}
\affiliation{Ames Laboratory and Department of Physics and Astronomy, Iowa State University, Ames, IA, 50011, USA}
\author{A.~I.~Goldman}
\affiliation{Ames Laboratory and Department of Physics and Astronomy, Iowa State University, Ames, IA, 50011, USA}
\author{R.~J.~McQueeney}\thanks{Corresponding author: mcqueeney@ameslab.gov}
\affiliation{Ames Laboratory and Department of Physics and Astronomy, Iowa State University, Ames, IA, 50011, USA}

\date{\today}

\begin{abstract}
Inelastic neutron scattering measurements on $\mathrm{Ba(Fe_{0.963}Ni_{0.037})_{2}As_{2}}$ manifest a neutron spin resonance in the superconducting state with anisotropic dispersion within the Fe layer. Whereas the resonance is sharply peaked at $\textbf{Q}_\mathrm{AFM}$ along the orthorhombic \textbf{a} axis, the resonance disperses upwards away from $\textbf{Q}_\mathrm{AFM}$ along the \textbf{b} axis. In contrast to the downward dispersing resonance and hour-glass shape of the spin excitations in superconducting cuprates, the resonance in electron-doped BaFe$_2$As$_2$ compounds possesses a magnon-like upwards dispersion.
\end{abstract}


\maketitle

A spin resonance, observed in inelastic neutron scattering measurements, appears in superconducting materials that do not possess a simple $s$-wave gap symmetry.\cite{Eschrig06,Maier08} Consequently, the spin resonance is considered to be a hallmark of unconventional superconductivity and highlights the important relationship between antiferromagnetic spin fluctuations and superconductivity.
The spin resonance has an intimate connection to the superconducting state.\cite{Eschrig06,Maier08, Bourges00,Pailhes04,Mazin08}  The inelastic neutron scattering (INS) signature of the resonance is a gapping of normal state spin fluctuations at the antiferromagnetic (AFM) wavevector, $\textbf{Q}_\mathrm{AFM}$, for energies well below the superconducting gap $2\Delta$, and an enhancement of magnetic spectral weight at an energy $\Omega_{0}<2\Delta$. The energy of the resonance at $\textbf{Q}_\mathrm{AFM}$ is found to be proportional to the superconducting (SC) transition temperature $T_\mathrm{c}$  ($\Omega_{0}/k_\mathrm{B}T_\mathrm{c} \approx 4-6 $) or 2$\Delta$ ($\Omega_{0}/2\Delta \approx 0.6 $) for a variety of unconventional superconductors.\cite{Yu09,Inosov11} The dispersion of the spin resonance, $\Omega_\mathbf{q}$ (where $\textbf{q}=\mathbf{Q}-\mathbf{Q}_\mathrm{AFM}$), is a measure of the resonance energy away from $\mathbf{Q}_\mathrm{AFM}$. 

\begin{figure}
\centering
\includegraphics[width=0.8\linewidth]{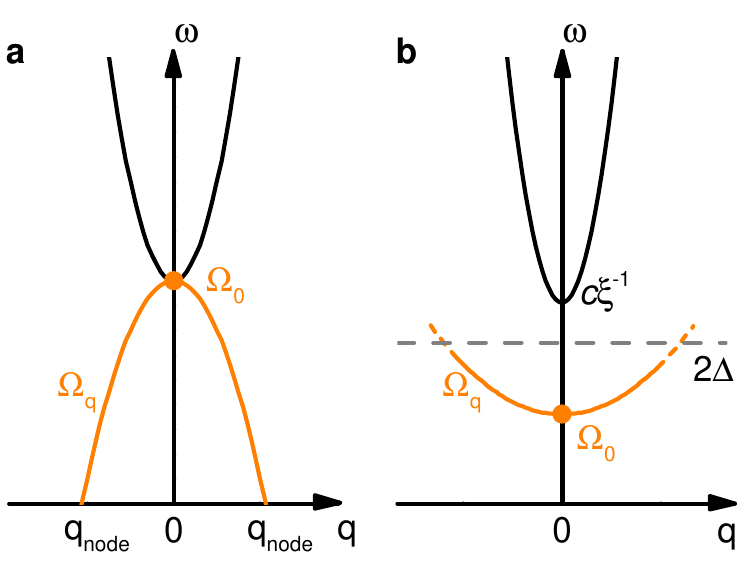}
\caption{ (a) \rm Spin fluctuations in the cuprates showing the hour-glass shape of the spin excitations.  The downward dispersing part (orange line) is the resonance mode for a $d$-wave gap and the upward part (black line) is the competing AFM spin fluctuations. (b) Spin fluctuations in our Ni-doped iron arsenide superconductor where both the resonance and competing AFM fluctuations disperse upwards.}
\label{fig1}
\end{figure}

In the case of the single-band cuprates, the dispersion of spin fluctuations forms a characteristic hour-glass shape below $T_\mathrm{c}$ where the resonance appears at the neck of the hour-glass with energy $\Omega_{0}$, as shown in Fig.~\ref{fig1}a.\cite{Bourges00,Pailhes04}  The downward dispersion ($\omega < \Omega_{0}$) is associated with the resonance itself, $\Omega_\mathbf{q}$, whereas the upward dispersion ($\omega > \Omega_{0}$) arises from normal state spin waves.\cite{Das12} In the cuprates, the downward resonance dispersion is determined by nodes in the $d$-wave SC gap at $\mathbf{q}=\mathbf{q}_\mathrm{node}$ where $\Delta_{\mathbf{q}_\mathrm{node}}=0$. 

Although the spin resonance has been clearly observed at $\textbf{Q}_\mathrm{AFM}$ in multi-band iron arsenide compounds studied to date, no dispersion of the resonance within the Fe layers (in the \textbf{ab} plane) has yet been reported and, complicating matters, band structure calculations predict both an upward~\cite{Zhang10} and downward~\cite{Maier12} dispersion depending on the details of the bands and the symmetry of the superconducting order parameter. In this Letter, we use inelastic neutron scattering to show that the resonance in Ba(Fe$_{0.963}$Ni$_{0.037}$)$_{2}$As$_{2}$ ($T_\mathrm{c}=$ \mbox{17 K}) disperses upwards. Using a simple mean-field approach with the assumption of $s^{\pm}$ order parameter, we show that the details of the dispersion, such as the resonance velocity, are determined by the normal state spin fluctuations.

The sample studied is a single crystal of $\mathrm{Ba(Fe_{1-x}Ni_{x})_{2}As_{2}}$ with $x=0.037$ weighing 436~mg. Rather than using co-aligned single crystals with a larger total mass, we chose to measure a single specimen of high crystallinity (mosaic width $<$ 0.44$^{\circ}$) to minimize effects of sample mosaic on the $\textbf{q}$-dependence of the spin fluctuations. Our preliminary neutron and x-ray scattering measurements show that the sample orders into an incommensurate AFM structure\cite{Kim12} below $T_\mathrm{N}=26$~K with a tetragonal-orthorhombic transition below $T_\mathrm{S}=29$~K.  The superconducting transition temperature is $T_\mathrm{c}=17$~K.  Additional details of crystal growth and characterization are described elsewhere.\cite{Canfield09} The sample was mounted in a closed-cycle refrigerator for temperature dependent studies on the HB3 neutron spectrometer at the High Flux Isotope Reactor at Oak Ridge National Laboratory using pyrolitic graphite (PG) (0,~0,~2) reflections for both monochromator and analyzer  and a configuration with $E_{f}=14.7$~meV, $48'-60'-80'-120'$ collimation and two PG filters before the analyzer.  We define $\mathbf{Q} = \left(\hh,\kk,L\right) = \frac{2\pi}{a}\hh\hat{\imath} + \frac{2\pi}{a}\kk\hat{\jmath} + \frac{2\pi}{c}L\hat{k}$ where the orthorhombic lattice constants, $a \approx b=5.6 $\,\AA~and $c=13 $\,\AA~were determined at $T=15$\,K.  The crystal was aligned in the $(1,0,1)-(0,1,0)$ scattering plane. Raw data were converted into the imaginary part of the magnetic susceptibility, $\chi''(\textbf{Q},\omega)$ after subtracting estimates of the non-magnetic background and correcting for the Bose population factor.  In addition, we performed inelastic neutron scattering measurements on the Cold Neutron Chopper Spectrometer (CNCS) at the Spallation Neutron Source at Oak Ridge National Laboratory using an incident energy of 10~meV.  Data were accumulated in the $[H,K,1]$ plane by rotating the crystal around a vertical (0,~1,~0)-axis and performing a series of exposures.

\begin{figure}
\includegraphics[width=0.8\linewidth]{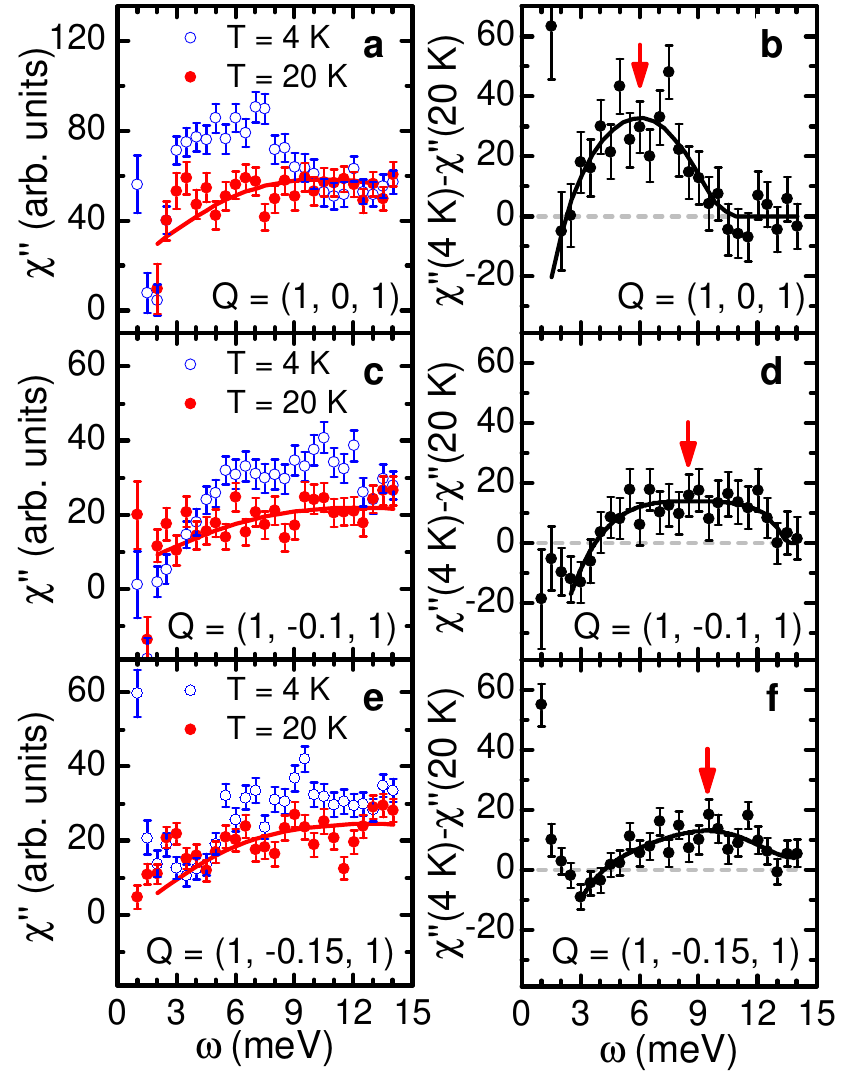}
\caption{ \rm The imaginary part of the magnetic susceptibility measured on HB3 at ({a}) $\textbf{Q}_\mathrm{AFM}=(1, 0, 1)$, ({c}) $\textbf{Q}=(1, -0.1, 1)$, and ({e}) $\textbf{Q}=(1, -0.15, 1)$ in the normal state at $T=20$~K (filled circles) and the superconducting state at $T=4$~K (open circles).  Red lines are model calculations of the normal state susceptibility assuming the NAFL spin fluctuations, as described in the text. The difference between superconducting and normal state susceptibilities is shown at ({b}) $\textbf{Q}_\mathrm{AFM}$, ({d}) $\textbf{Q}=(1, -0.1, 1)$, and ({f}) $\textbf{Q}=(1, -0.15, 1)$. Black lines are guide to the eyes and arrows indicate the resonance peak obtained from Gaussian fit.}
\label{fig3}
\end{figure}

Figure~\ref{fig3}a shows the energy dependence of the imaginary part of the magnetic susceptibility, $\chi''(\textbf{Q}_\mathrm{AFM},\omega)$, for $\mathrm{Ba(Fe_{0.963}Ni_{0.037})_{2}As_{2}}$ in the normal state at $T=20$~K and the superconducting state at $T=4$~K, where $\textbf{Q}_\mathrm{AFM} = (1, 0, 1)$ in orthorhombic coordinates. The neutron spin resonance peaks at an energy of $\Omega_{0} \approx $ 6 meV, as made clearer by the difference plot in Fig.~\ref{fig3}b showing the resonance enhancement. 
At $\textbf{Q}_\mathrm{AFM}$, the resonance extends up to approximately 10~meV, above which the susceptibilities of the normal and SC states are equivalent.  Measurements of the energy spectrum at positions offset from $\textbf{Q}_\mathrm{AFM}$ in the transverse direction [$\textbf{Q}=(1, -0.1, 1)$ and $(1, -0.15, 1)$] show that the center of the resonance intensity shifts up to higher energy (Figs.~\ref{fig3}c$-$f). In addition, we observe that the resonance spectral weight extends to at least 14~meV.  Thus, despite the very broad lineshapes, measurements of the susceptibility away from $\textbf{Q}_\mathrm{AFM}$ show that the spectral weight of the resonance has moved to higher energy, i.e. \textit{the resonance is dispersing upwards, unlike the cuprates}.

\begin{figure*}[t!]
\includegraphics[width=0.85\linewidth]{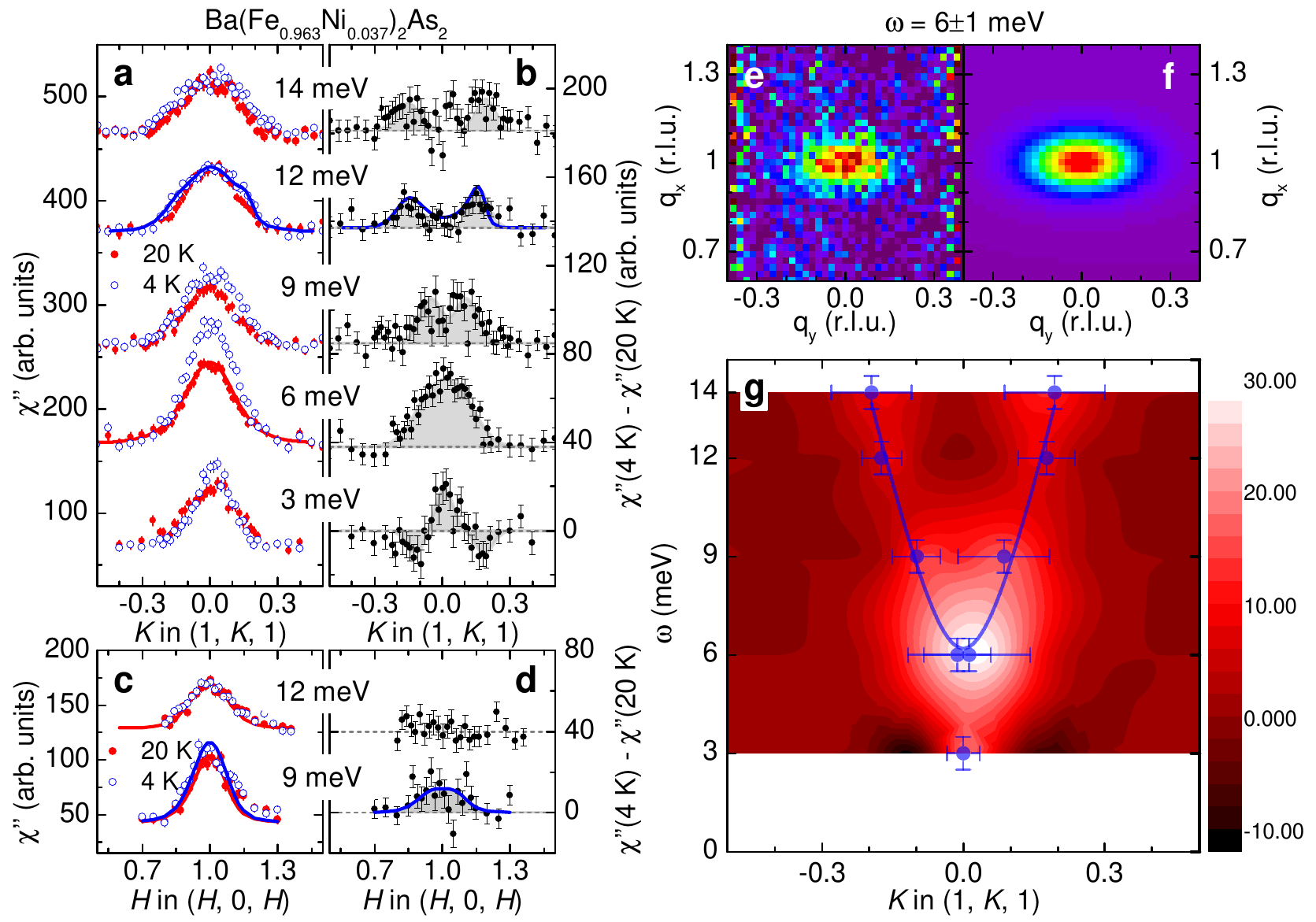}
\caption{({a}) \rm The imaginary part of the magnetic susceptibility measured on HB3 at various energy transfers along the [0, $\kk$, 0]-direction transverse to $\textbf{Q}_\mathrm{AFM}=(1, 0, 1)$ in the normal state at $T=$ 20~K (filled circles) and the superconducting state at $T=4$~K (open circles). ({b}) The difference of superconducting minus normal state susceptibilities from ({a}). ({c}) The susceptibilities along the [\hh, 0, \hh]-direction longitudinal to  $\textbf{Q}_\mathrm{AFM}$ in the normal (filled circles) and superconducting states (open circles) and (d) the difference. Red and blue lines are model calculations of the normal state susceptibility and superconducting resonance, respectively, as described in the text. ({e}) CNCS measurements of the normal state spin fluctuations in the ($\hh$, $\kk$, 1) plane at $\omega =6$~meV and $T=25$~K. ({f}) Fit of the CNCS data to a model of NAFL, as described in the text. ({g}) Contour plot showing the difference of the magnetic susceptibilities in the superconducting and normal states measured on HB3 depicting the dispersion of the resonance in the transverse direction.  Symbols correspond to fitted values of the resonance maxima and the line is a fit to the magnonlike dispersion given in Eq.~\ref{eq13}. The vertical error bars indicate the energy resolution of HB3 and horizontal error bars indicate the full-width-half-maximum of fits.}
\label{fig4}
\end{figure*}

The observation of resonance dispersion is confirmed by a series of constant-energy $\textbf{Q}$-scans in the normal and SC states in the vicinity of $\textbf{Q}_\mathrm{AFM}=(1, 0, 1)$, focusing on the transverse direction $\textbf{Q}=(1, \kk, 1)$ [$\textbf{q}=(0, \kk, 0)$] as shown in Figs.~\ref{fig4}a and b. Here, the resonance is clearly identified by the enhanced susceptibility at $\textbf{Q}_\mathrm{AFM}$ and $\omega \approx \Omega_{0} =6$~meV. As the energy is increased, the spectral weight progressively moves away from $\textbf{Q}_\mathrm{AFM}$ and weakens. Plots of the difference between the susceptibility in the normal and superconducting states more clearly show that the resonance, $\Omega_{\textbf{q}}$, disperses upwards away from $\textbf{Q}_\mathrm{AFM}$ (Fig.~\ref{fig4}b).
In contrast, along the longitudinal $(\hh, 0, \hh)$ direction in our scattering geometry, we find that only a weak resonance enhancement remains at $\textbf{Q}_\mathrm{AFM}$ at 9~meV and no resonance enhancement is observed at 12~meV, i.e. there is surprisingly no indication of dispersion in the longitudinal direction (Figs.~\ref{fig4}e and f).  This observation is consistent with early measurements of the resonance in both Ni (ref.~\onlinecite{Chi09}) and Co-doped (ref.~\onlinecite{Lumsden09,Pratt09,Christianson09}) BaFe$_2$As$_2$ that find a sharply defined resonance in the $\hh$ (orthorhombic \textbf{a}) direction. Thus, while the resonance dispersion is observed in the transverse direction, it is apparently too steep to observe a splitting in the longitudinal direction due to the limited instrumental resolution (we will return to this point below). As discussed below, the $\mathbf{Q}$-space anisotropy of the spin resonance in the SC state may be associated with the normal state spin fluctuations within the Fe layers that possess a two-fold (elliptical) anisotropy (Fig.~\ref{fig4}e).\cite{Diallo10,Li10,Tucker12, Park10}

We now turn to a discussion of the spin dynamics in the normal and SC states, which leads to a straightforward interpretation of the resonance dispersion and its anisotropy. Despite the presence of weak AFM ordering, the normal state spin dynamics of the underdoped BaFe$_{2}$As$_{2}$ compounds at low energies can be understood using the model of nearly AFM Fermi liquids (NAFL).\cite{Inosov09, Diallo10} The imaginary part of the dynamical susceptibility, as measured by INS, can be written as in ref.~\onlinecite{Inosov09}, 
\begin{equation}
\chi_{n}''(\mathbf{q},\omega) = \frac{\chi_{0}\xi_{\textbf{q}}^{2}\Gamma_{\textbf{q}}\omega}{\omega^{2}+\Gamma_{\textbf{q}}^{2}\lbrack 1+\xi_{\textbf{q}}^{2}q^{2}\rbrack^{2}}
\label{eq3}
\end{equation}
where $\xi_{\textbf{q}}$ is the AFM correlation length, $\Gamma_{\textbf{q}}$ is the relaxation width due to the decay of spin waves into electron-hole pairs (Landau damping), and the subscript $\textbf{q}$ allows for anisotropy in these quantities.

The anisotropy of the normal state susceptibility has been studied in detail.~\cite{Tucker12} The two-fold in-plane anisotropy of the magnetic correlation length is most important for the subsequent discussion,
\begin{equation}
\xi_{\textbf{q}}^{2}q^{2} = \xi^{2}[q^{2}+\eta(q_{x}^2-q_{y}^2)]
\label{eq8}
\end{equation}
and we assume in Eq.~\ref{eq3} that the damping $ \Gamma_{\textbf{q}} \approx \Gamma_{0}$ is isotropic and constant throughout the Brillouin zone. Since the resonance dispersion observed along the $L$ (\textbf{c}) direction in $A$Fe$_2$As$_2$ ($A$ = Ca, Sr, and Ba) based superconductors is weak, we assume $q_z$ gives a negligible effect. Typical fits to the normal state data in terms of the NAFL convolved with the instrumental resolution are shown in Figs.~\ref{fig3},~\ref{fig4}a,~\ref{fig4}c, and~\ref{fig4}f with $\Gamma_{0}$ fixed to 10 meV.~\cite{Tucker12}  We obtain the following parameters; $\xi=9.5(4)$ \AA\ and $\eta=0.5(1)$ corresponding to different correlation lengths $11.8(7)$~\AA\ and $6.5(3)$~\AA\ in the $\hh$ (orthorhombic \textbf{a}) and $\kk$ (\textbf{b}) directions within the Fe layers, respectively.

The neutron spin resonance is reasonably well understood as an excitonic bound state in an itinerant antiferromagnet.\cite{Eschrig06,Maier08}  In this picture, fermions form singlet Cooper pairs and electron-hole (singlet-triplet) excitations appear with a threshold energy $|\Delta_{\textbf{k}}|+|\Delta_{\textbf{k}+\textbf{Q}_\mathrm{AFM}}|$.  These electron-hole excitations will form a bound-state inside the SC gap (an exciton) as a consequence of interactions already present in the normal state.  The spectral features of the bound-state can be understood within a mean-field or random-phase approximation (RPA) approach to the interactions, as described below.

Following closely the arguments from Ref.~\onlinecite{Chubukov01}, the RPA analysis of the susceptibility in the superconducting state leads to an equation for the dispersion of the resonance mode,
\begin{equation}
\Omega_{\textbf{q}}^{2} = \Delta_{\textbf{q}}\Gamma_{\textbf{q}}(1+\xi_{\textbf{q}}^{2}q^{2}).
\label{eq12}
\end{equation}
where $2\Delta_{\textbf{q}}=|\Delta_{\textbf{k}}|+|\Delta_{\textbf{k}+\textbf{Q}_\mathrm{AFM}+\textbf{q}}|$ is the fermion gap at two points on the Fermi surface connected by $\textbf{q}+\textbf{Q}_\mathrm{AFM}$ and   $\Omega_{0}=\sqrt{\Delta_{0}\Gamma_{0}}$ is the resonance energy at $\textbf{Q}_\mathrm{AFM}$ ($\mathbf{q}=0$).  Note that for a single-band model with a $d$-wave gap possessing nodes at $\mathbf{q}_\mathrm{node}$, $\Delta_{0}$ is a maximum and $\Delta_{\textbf{q}_\mathrm{node}} \rightarrow 0$ at points on the Fermi surface, thereby forcing $\Omega_{\textbf{q}_\mathrm{node}} \rightarrow 0$ and resulting in the downward dispersion shown in Fig.~\ref{fig1}a. The simple RPA model therefore captures the essential features resulting in the downward dispersion in the cuprates.

Although the \textbf{q}-dependence of the resonance in Eq.~\ref{eq12} has several contributions (such as \textbf{q}-dependent gap or damping effects), here we choose to apply the RPA approach for the iron arsenides under the assumption that the \textbf{q}-dependence arises only from anisotropy in the normal state spin fluctuations.  Assuming the proposed $s^{\pm}$ gap symmetry for our compound, $\Delta_{\textbf{q}}=\Delta_{0}$ for all $\textbf{q}$ and the equation above takes the form of a gapped magnon with upward dispersion
\begin{equation}
\Omega_{\textbf{q}} = \sqrt{\Omega_{0}^{2}+c_{res,\textbf{q}}^{2}q^{2}}
\label{eq13}
\end{equation}
where $c_{res,\textbf{q}}^{2} = \Omega_{0}^{2} \xi_{\mathbf{q}}^{2}$ is the velocity of the resonance mode which becomes anisotropic itself due to the anisotropy in the normal state spin-spin correlation length. More detailed four-band RPA calculations (ref.~\onlinecite{Maiti11}) and two-orbital calculations of the susceptibility within the self-consistent fluctuation exchange (FLEX) approximation (ref.~\onlinecite{Zhang10}) arrive at similar conclusions regarding the upward resonance dispersion, at least at low doping concentrations. With increased electron doping, other multi-band RPA calculations have predicted the development of an incommensurate normal state response.\cite{Park10,Maiti11,Maier12} In the superconducting state, these calculations predict that the resonance itself is also incommensurate and first disperses downwards to the incommensurate wavevector, and then upwards. Our observation highlighting the upward dispersing and commensurate response is, therefore, consistent with the multi-band RPA approach for underdoped compositions. We note that an incommensurate resonance was reported for Ba(Fe$_{0.925}$Ni$_{0.075}$)$_2$As$_2$ by Luo $et~al.$\cite{Luo12} at a higher composition than reported here. This evolution towards an incommensurate response would be consistent with the general trend outlined above. However, the incommensurate splitting was only observed in the superconducting state (not in the normal state) and found to disperse upwards away from \textbf{Q}$_\mathrm{AFM}$ rather than downwards.

We predict that the anisotropic resonance velocities, $\sqrt{\Omega_{0}^{2}\xi^{2}(1 \pm \eta)}$, in the Fe layers to be $85(5)$~meV~\AA\ and $50(5)$~meV~\AA\ in the $\hh$ and $\kk$ directions, respectively, using values ($\xi=9.5(4)$~\AA\ and $\eta=0.5(1)$) obtained from the normal state spin fluctuations. These predictions can be compared to fits made to the total susceptibility ($\chi''=\chi''_{n}+\chi''_{res}$) after convolution with the instrumental resolution function, where $\chi''_{res}$ is the resonance enhancement that obeys the dispersion relation in Eq.~\ref{eq13} and $\chi''_{n}$ is given by Eq.~\ref{eq3}.  Typical fits are shown as the blue line in Figs.~\ref{fig4}a and b.  A contour plot of $\chi''_{res}$ obtained from INS data along with fitted values of the peak positions is shown in Fig.~\ref{fig4}g and we arrive at $63(2)$~meV~\AA\ for the observed resonance velocity in the $\kk$ direction, in good agreement with the simple analysis above. 

The resonance velocity is to be contrasted with the velocity of the normal state AFM spin waves which are much steeper $c \approx$ 450~meV~\AA\ $\gg$ $c_{res}$. Within the disordered AFM state model, dispersive features of the normal state AFM spin waves will appear only at much higher energies, $\omega > c\xi^{-1} \approx$ 60~meV, as shown in Fig.~\ref{fig1}b. This disparity in the magnitude of the normal state and resonance velocities eliminates the possibility that resonance may be interpreted as damped AFM spin waves that sharpen up in the SC state (see Supplement for more information). We also note that limitations due to experimental resolution may prevent observation of the splitting in the longitudinal direction with the predicted velocity of $85(5)$~meV~\AA\, as shown in Fig.~\ref{fig4}c and d.

In conclusion, we find that the dispersion of the neutron spin resonance in $\mathrm{Ba(Fe_{0.963}Ni_{0.037})_{2}As_{2}}$ can be interpreted based on the assumption of extended $s$-wave superconductivity and the properties of the NAFL spin fluctuations in the normal state, most notably the anisotropic spin-spin correlation length.  Such observations strongly support an excitonic picture for the spin resonance.  Similar reasoning may also explain the weak resonance dispersion observed along the direction perpendicular to the layers ($L$-direction) in $A$Fe$_2$As$_2$ based superconductors. For quasi-two-dimensional normal state spin fluctuations observed in optimally doped samples,  the resonance will have no dispersion along $L$ (appear flat) due to the vanishing correlation length.  However, underdoped samples with long-range AFM order, such as Ba(Fe$_{1-x}$Co$_{x}$)$_{2}$As$_{2}$ with $x=$ 0.04 \cite{Christianson09} and 0.047\cite{Pratt09,Pratt10}, still possess weak interlayer spin correlations. Correspondingly a finite, albeit small, $L$-dispersion is observed in these samples (with a velocity of $\sim 6$~meV~\AA).  In a similar vein, doping will also affect both the interlayer and intralayer correlation length and consequently the resonance dispersion is expected to be composition dependent.  This could explain recent observations of a large transverse resonance splitting in overdoped Ba(Fe$_{1-x}$Ni$_{x}$)$_2$As$_2$.\cite{Luo12}

\begin{acknowledgments} 
The work at Ames Laboratory was supported by the U.S. Department of Energy, Office of Basic Energy Science, Division of Materials Sciences and Engineering under Contract No. DE-AC02-07CH11358. Work at Oak Ridge National Laboratory is supported by U.S. Department of Energy, Office of Basic Energy Sciences, Scientific User Facilities Division.
\end{acknowledgments}

\bibliographystyle{apsrev}
\bibliography{ni_ins}

\end{document}


\title{Supplement: Magnon-like dispersion of neutron spin resonance in Ni-doped BaFe$_{2}$As$_2$}

\author{M.~G.~Kim}\thanks{Current electronic address: mgkim@lbl.gov}
\affiliation{Ames Laboratory and Department of Physics and Astronomy, Iowa State University, Ames, IA, 50011, USA}
\author{G.~S.~Tucker}
\affiliation{Ames Laboratory and Department of Physics and Astronomy, Iowa State University, Ames, IA, 50011, USA}
\author{D.~K.~Pratt}
\affiliation{Ames Laboratory and Department of Physics and Astronomy, Iowa State University, Ames, IA, 50011, USA}
\author{S.~Ran}
\affiliation{Ames Laboratory and Department of Physics and Astronomy, Iowa State University, Ames, IA, 50011, USA}
\author{A.~Thaler}
\affiliation{Ames Laboratory and Department of Physics and Astronomy, Iowa State University, Ames, IA, 50011, USA}
\author{A.~D.~Christianson}
\affiliation{Oak Ridge National Laboratory, Oak Ridge, TN, 37831, USA}
\author{K.~Marty}
\affiliation{Oak Ridge National Laboratory, Oak Ridge, TN, 37831, USA}
\author{S.~Calder}
\affiliation{Oak Ridge National Laboratory, Oak Ridge, TN, 37831, USA}
\author{A.~Podlesnyak}
\affiliation{Oak Ridge National Laboratory, Oak Ridge, TN, 37831, USA}
\author{S.~L.~Bud'ko}
\affiliation{Ames Laboratory and Department of Physics and Astronomy, Iowa State University, Ames, IA, 50011, USA}
\author{P.~C.~Canfield}
\affiliation{Ames Laboratory and Department of Physics and Astronomy, Iowa State University, Ames, IA, 50011, USA}
\author{A.~Kreyssig}
\affiliation{Ames Laboratory and Department of Physics and Astronomy, Iowa State University, Ames, IA, 50011, USA}
\author{A.~I.~Goldman}
\affiliation{Ames Laboratory and Department of Physics and Astronomy, Iowa State University, Ames, IA, 50011, USA}
\author{R.~J.~McQueeney}\thanks{Corresponding author: mcqueeney@ameslab.gov}
\affiliation{Ames Laboratory and Department of Physics and Astronomy, Iowa State University, Ames, IA, 50011, USA}

\maketitle

The following supplemental information covers the derivation of the dynamical magnetic susceptibility of itinerant antiferromagnets in the normal and superconducting states and the approximations made in the derivation.

\vspace{2 mm}\noindent \textit{Normal state}.  Despite the presence of weak AFM ordering, the normal state spin dynamics in the underdoped BaFe$_{2}$As$_{2}$ compounds can be understood using a model of spin waves in a disordered AFM state.  In the long-wavelength limit, the dynamical susceptibility can be written as~\cite{Chubukov01}
\begin{equation}
\chi_{n}(\mathbf{q},\omega) = \chi_{0}\lbrack\xi_{\textbf{q}}^{-2} + q^{2}-\frac{\omega^{2}}{c_{\textbf{q}}^{2}}-\Pi_{n}(\textbf{q},\omega)\rbrack^{-1}
\label{eq1}
\end{equation}
where $\xi_{\textbf{q}}$ is the AFM correlation length, $c_{\textbf{q}}$ is the AFM spin wave velocity and the subscript $\textbf{q}$ allows for anisotropy in these quantities.  The wavevector $\textbf{q}=\textbf{Q}-\textbf{Q}_\mathrm{AFM}$ is measured relative to the antiferromagnetic wavevector $\textbf{Q}_\mathrm{AFM}$.  For a spin-density wave state, $\Pi_{n}(\textbf{q},\omega) = i\gamma_{\textbf{q}}\omega$ is purely imaginary and arises from the decay of spin waves into electron-hole pairs (Landau damping). Neutron scattering measures the imaginary part of the susceptibility, given by
\begin{equation}
\chi_{n}''(\mathbf{q},\omega) = \frac{\chi_{0}\xi_{\textbf{q}}^{2}\Gamma_{\textbf{q}}\omega}{\omega^{2}+\Gamma_{\textbf{q}}^{2}\lbrack 1+\xi_{\textbf{q}}^{2}q^{2}-\omega^{2}\xi_{\textbf{q}}^{2}/c_{\textbf{q}}^{2}\rbrack^{2}}
\label{eq2}
\end{equation}
where $\Gamma_{\textbf{q}}^{-1}=\gamma_{\textbf{q}}\xi_{\textbf{q}}^{2}$ is the relaxation width due to Landau damping.  We have showed previously in underdoped Ba(Fe$_{1-x}$Co$_{x}$)$_{2}$As$_{2}$ that on average $c \approx $ 450 meV\AA\ and $\xi \approx$ 7.5 \AA\  and $c\xi^{-1} \approx$ 60 meV.\cite{Tucker12}  So for energies relevant to the superconducting resonance $\omega < 2\Delta \approx $ 10 meV $\ll c\xi^{-1}$, we can ignore the dispersive spin waves and the normal state susceptibility falls into the diffusive limit
\begin{equation}
\chi_{n}''(\mathbf{q},\omega) = \frac{\chi_{0}\xi_{\textbf{q}}^{2}\Gamma_{\textbf{q}}\omega}{\omega^{2}+\Gamma_{\textbf{q}}^{2}\lbrack 1+\xi_{\textbf{q}}^{2}q^{2}\rbrack^{2}}.
\label{eq3}
\end{equation}

Fig. S1(a) shows a typical form of the response of an itinerant antiferromagnet in the normal state.

\renewcommand{\thefigure}{S\arabic{figure}}
\renewcommand{\figurename}{\bfseries Figure}
\begin{figure}[t]
\includegraphics[width=0.70\linewidth]{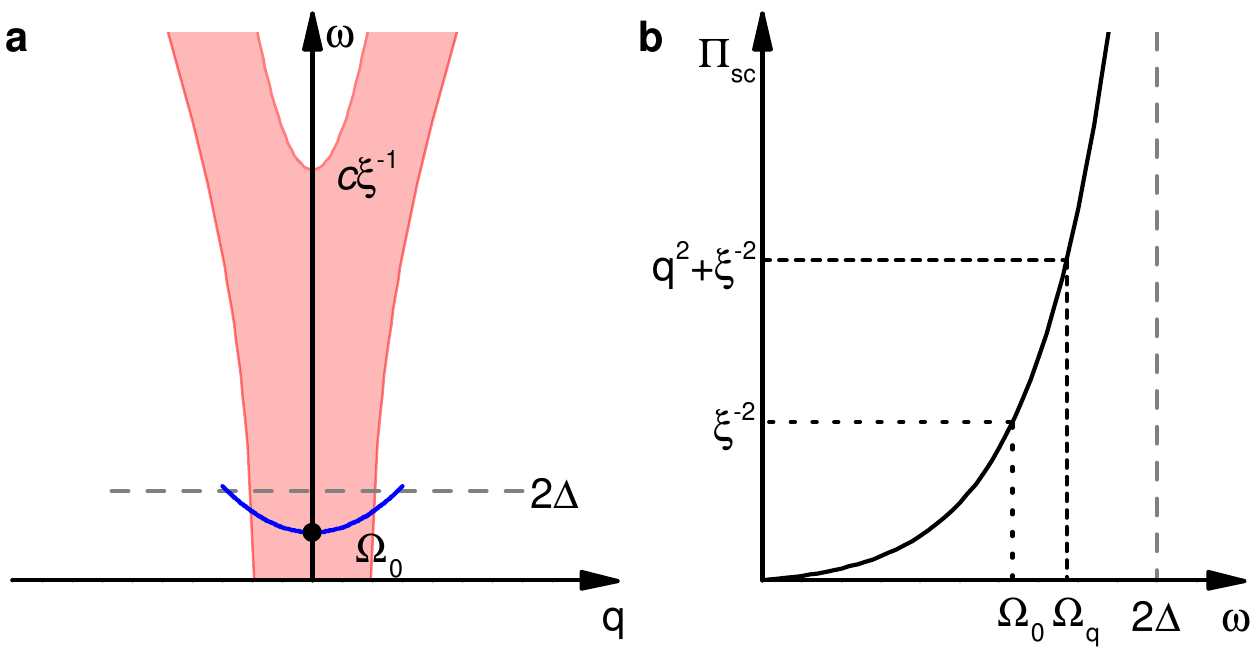}
\caption{\footnotesize  {\bf The response of an itinerant antiferromagnet and  the resonance dispersion for an s-wave superconductor} \rm \textbf{(a)} A typical form of the response of an itinerant antiferromagnet in the normal (red shade). The superconducting response is indicated by the resonance dispersion (blue line). Dispersion of the normal state AFM spin waves appear above $\omega \gtrsim c\xi^{-1}$ and the SC resonance occurs at $\Omega_0 \ll c\xi^{-1}$ and disperses upwards. \textbf{(b)} A graphical explanation of the resonance dispersion for an s-wave superconductor. The solid line corresponds to the quadratic behavior of $\Pi_\mathrm{SC}$ in the superconducting state. The SC resonance appears at $\omega = \Omega_0$ when $\Pi_\mathrm{SC} = \xi^{-2}$ (with $\mathbf{q}^2=0$). At $\mathbf{q} \neq 0$, the resonance will always occur at a larger energy, $\Omega_\mathrm{q}$.}
\label{fig1}
\end{figure}

\vspace{2 mm}\noindent \textit{Superconducting state}. In the superconducting state,  a similar form for the susceptibility can be used \cite{Chubukov01}
\begin{equation}
\chi_{sc}(\mathbf{q},\omega) = \chi_{0}\lbrack\xi_{\textbf{q}}^{-2} + q^{2}-\frac{\omega^{2}}{c_{\textbf{q}}^{2}}-\Pi_{sc}(\textbf{q},\omega)\rbrack^{-1}.
\label{eq4}
\end{equation}
However, in this case the opening of the superconducting gap eliminates the imaginary contribution of the Landau damping to the self energy.  Instead, the smallest non-vanishing term is quadratic in energy, $\Pi_{sc}(\textbf{q},\omega) \sim \gamma_{\textbf{q}}\omega^{2}/\Delta_{\textbf{q}}$.  Here $2\Delta_{\textbf{q}}=|\Delta_{\textbf{k}}|+|\Delta_{\textbf{k}+\textbf{Q}_\mathrm{AFM}+\textbf{q}}|$ is the fermion gap at two points on the Fermi surface connected by $\textbf{q}+\textbf{Q}_\mathrm{AFM}$. The susceptibility can then be written 
\begin{equation}
\chi_{sc}(\mathbf{q},\omega) \sim \chi_{0}\lbrack\xi_{\textbf{q}}^{-2} + q^{2}-\omega^{2}(\frac{1}{c_{\textbf{q}}^{2}}+\frac{\gamma_{\textbf{q}}}{\Delta_{\textbf{q}}})\rbrack^{-1}.
\label{eq4}
\end{equation}
Again using the properties of the normal state spin fluctuations (the isotropic damping, $\Gamma_0$, and the AFM correlation length, $\xi$)~\cite{Tucker12} and the superconducting gap magnitude ($\Delta_0$), we estimate that  $\sqrt{\Delta_{0}\Gamma_{0}\xi^{2}} \approx \sqrt{(5\ \text{meV})(10\ \text{meV})(8\ \text{\AA} )^{2}} \approx$ 60 meV\AA.  As this is much smaller than the the AFM spin wave velocity, this shows that we can again ignore the spin wave term and we arrive at the following approximation for the susceptibility:
\begin{equation}
\chi_{sc}(\mathbf{q},\omega) \sim \chi_{0}\xi_{\textbf{q}}^{2}\lbrack 1 + \xi_{\textbf{q}}^{2}q^{2}-\frac{\omega^{2}}{\Delta_{\textbf{q}}\Gamma_{\textbf{q}}}\rbrack^{-1}.
\label{eq5}
\end{equation}
While this form of the susceptibility contains only a real-part and is not directly comparable to the neutron scattering data, the resonance condition will correspond to the vanishing of the term in square brackets in the denominator, leading to an equation for the dispersion of the resonance mode
\begin{equation}
\Omega_{\textbf{q}}^{2} = \Delta_{\textbf{q}}\Gamma_{\textbf{q}}(1+\xi_{\textbf{q}}^{2}q^{2})
\label{eq6}
\end{equation}
where at $\textbf{Q}_\mathrm{AFM}$ ($\textbf{q}=0$), $\Omega_{0}=\sqrt{\Delta_{0}\Gamma_{0}}$ is the resonance energy.  Note that for a $d$-wave gap, $\Delta_{0}$ is a maximum and $\Delta_{\textbf{q}_\mathrm{node}} \rightarrow 0$ at the node wavevector $\textbf{q}_\mathrm{node}$ resulting in a downward dispersion.  For the isotropic $s$-wave gap of the iron arsenides, $\Delta_{\textbf{q}}=\Delta_{0}$ for all $\textbf{q}$ and the equation above takes the form of a gapped magnon 
\begin{equation}
\Omega_{\textbf{q}} = \sqrt{\Omega_{0}^{2}+c_{res,\textbf{q}}^{2}q^{2}}
\label{eq6}
\end{equation}
where $c_{res,\textbf{q}}^{2} = \Delta_{0}\Gamma_{\textbf{q}}\xi_{\textbf{q}}^{2}$ is the velocity of the resonance mode.  One can refer to Fig. S1(b) for a graphical explanation of the resonance dispersion for an $s$-wave superconductor.  In this figure, we see that the resonance condition corresponds to the vanishing of the susceptibility in Eq.~4 (or denominator in Eq.~6) where $q^2+\xi^{-2}-\Pi_\mathrm{SC}=0$. The resonance will always occur at a larger energy for non-zero values of $q$ when compared to $q=0$.

\vspace{2 mm}\noindent \textit{Anisotropy}.  The normal state spin fluctuations in the iron arsenides possess a two-fold anisotropy about $\textbf{Q}_\mathrm{AFM}$.  This anisotropy has been studied in detail \cite{Diallo10, Tucker12} and can be cast in the form of anisotropies in the magnetic correlation length and damping, 
\begin{equation}
\xi_{\textbf{q}}^{2}q^{2} = \xi^{2}[q^{2}+\eta_{\xi}(q_{x}^2-q_{y}^2)]
\label{eq4}
\end{equation}
and
\begin{equation}
\Gamma_{\textbf{q}} = \Gamma_{0}[1+\alpha^{2}(q^{2}+\eta_{\Gamma}(q_{x}^2-q_{y}^2)]
\label{eq8}
\end{equation}
where $\textbf{q}$ is defined in the coordinate system of the low temperature orthorhombic structure, $\eta_{\xi}$ is the anisotropy of the AFM correlation length, $\alpha$ is a constant allowing for momentum-dependent Landau damping, and $\eta_{\Gamma}$ is the anisotropy of the Landau damping.  This anisotropy of the normal state spin fluctuations will lead to anisotropy in the resonance dispersion transverse ($T$, $q=q_{y}$) and longitudinal ($L$, $q=q_{x}$) to $\textbf{Q}_\mathrm{AFM} = (1,0,1)$.  The resonance velocities in these two directions are

\begin{equation}
c_{res,T}^{2}= \Delta_{0}\Gamma_{0}[\xi^{2}(1-\eta_{\xi})+\alpha^{2}(1-\eta_{\Gamma})]
\label{eq9}
\end{equation}

and

\begin{equation}
c_{res,L}^{2}= \Delta_{0}\Gamma_{0}[\xi^{2}(1+\eta_{\xi})+\alpha^{2}(1+\eta_{\Gamma})]
\label{eq10}
\end{equation}

Referring to fits to the normal state spin fluctuations in a Co-substituted sample one last time\cite{Tucker12}, we have the following parameters; $\xi=$7.5 \AA\, $\eta_{\xi} =$ 0.6, $\Gamma_{0} =$ 11 meV, $\alpha =$ 2 \AA\, and $\eta_{\Gamma} =$ -1.  We notice that, for this particular set of parameters, the anisotropy in the damping contributes very little to the resonance velocity and can be ignored, resulting in the simple forms for the velocities.  
\begin{equation}
c_{res,T}^{2}= \Delta_{0}\Gamma_{0}\xi^{2}(1-\eta_{\xi})
\label{eq11}
\end{equation}
and
\begin{equation}
c_{res,L}^{2}= \Delta_{0}\Gamma_{0}\xi^{2}(1+\eta_{\xi})
\label{eq12}
\end{equation}

The work at Ames Laboratory was supported by the U.S. Department of Energy, Office of Basic Energy Science, Division of Materials Sciences and Engineering under Contract No. DE-AC02-07CH11358. Work at Oak Ridge National Laboratory is supported by U.S. Department of Energy, Office of Basic Energy Sciences, Scientific User Facilities Division.